%% file: bf0342.tex
\documentclass[aps,prl,preprint,tightenlines,superscriptaddress,showpacs,byrevtex]{revtex4}

\usepackage{graphicx} % Include figure files
\usepackage{dcolumn}  % Align table columns on decimal point

\graphicspath{{ps}}

\begin{document}

%\vspace*{-3\baselineskip}
\resizebox{!}{3cm}{\includegraphics{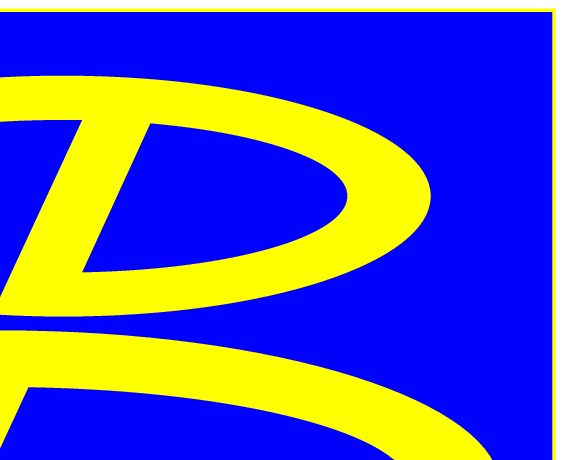}}

\preprint{\vbox{ \hbox{   }
                 \hbox{BELLE-CONF-0342}
		 %\hbox{Version2.3}
                 %\hbox{EPS Parallel Sessions: 3, 10, or 12}
                 %\hbox{EPS-ID nnn}
                 %\hbox{hep-ex nnnn, if available}
}}

%DRM \title{ \quad\\[0.5cm]  Study of time-dependent $CP$-violation parameters
%DRM in $B^0 \rightarrow J/\psi ~\pi^0$ decays}
\title{ \quad\\[0.5cm]  Study of time-dependent $CP$-violation 
in $B^0 \rightarrow J/\psi ~\pi^0$ decay}

%%%% insert the authorlist here. BEFORE the abstract !!!

\input author-conf2003.tex

%\collaboration{Belle Collaboration}
\noaffiliation

%\today
\begin{abstract}
We report a measurement of $CP$ asymmetry parameters in the 
%DRM $b \rightarrow c\bar{c}d$ transition-induced decay
$b \rightarrow c\bar{c}d$-transition-induced decays
%DRMof $B^0(\bar{B^0}) \rightarrow J/\psi ~\pi^0$ using
%of $B^0(\bar{B^0}) \rightarrow J/\psi ~\pi^0$.  
$B^0(\bar{B^0}) \rightarrow J/\psi ~\pi^0$.  
The analysis is based on 
a 140 fb$^{-1}$ data sample accumulated at
the $\Upsilon(4S)$ resonance
by the Belle detector 
%at the KEKB asymmetric energy $e^+ e^-$ collider.
at the KEKB asymmetric-energy $e^+ e^-$ collider.
%DMat the KEKB energy-asymmetric $e^+ e^-$ collider.
We fully reconstruct one neutral $B$ meson in the
%DRM $J/\psi ~\pi^0$ final state and the accompanying
$J/\psi ~\pi^0$ final state.   The accompanying
$B$ meson flavor is identified by its decay products.
From the distribution of proper time intervals between the two
$B$ decays, we obtain the following $CP$ violation parameters: 
%\begin{eqnarray}\nonumber
%{\cal S}_{J/\psi \pi^0}
%&=& -0.72 \displaystyle{^{+0.42}_{-0.37}}(\mbox{stat})
%\pm x.xx(\mbox{syst}) \\
%{\cal A}_{J/\psi \pi^0}
%&=& -0.01 \displaystyle{^{+0.29}_{-0.28}}(\mbox{stat})
%\pm x.xx(\mbox{syst}).  \nonumber
%\end{eqnarray}
\begin{eqnarray}\nonumber
{\cal S}_{J/\psi \pi^0}
&=& -0.72 \pm0.42 (\mbox{stat}) \pm 0.08(\mbox{syst}) \\
{\cal A}_{J/\psi \pi^0}
&=& -0.01 \pm0.29 (\mbox{stat}) \pm 0.07(\mbox{syst}).  \nonumber
\end{eqnarray}
\end{abstract}

%\pacs{13.65.+i, 13.25.Gv, 14.40.Gx}

\maketitle

%%%% keep the final version single-spaced
\tighten

{\renewcommand{\thefootnote}{\fnsymbol{footnote}}}
\setcounter{footnote}{0}

%%%%%%%
\section{Introduction}
%%%%%%
In the standard model (SM), the Kobayashi-Maskawa (KM)
quark-mixing matrix\cite{KM}
%DRM has an irreducible complex phase which causes $CP$ violation in weak
has an irreducible complex phase that gives rise to $CP$ violation in weak
interactions.
In particular, the SM predicts large $CP$-violating asymmetries in the
time-dependent rates of $B^0$ and $\bar{B^0}$ decays into a common
$CP$ eigenstate $f_{CP}$\cite{sanda}.
In the decay chain $\Upsilon(4S)\rightarrow B^0 \bar{B}^0
\rightarrow f_{CP}f_{\rm tag}$,
where one of the $B$ mesons decays at time $t_{CP}$ to a final state
$f_{CP}$
and the other decays at time $t_{\rm tag}$ to a final state
$f_{\rm tag}$ that distinguishes between $B^0$ and $\bar{B}^0$,
the decay rate has a time dependence
given by~\cite{CPVrev}
\begin{equation}
\label{eq:psig}
{\cal P}(\Delta t) =
\frac{e^{-|\Delta_{t}|/{\tau_{B^0}}}}{4{\tau_{B^0}}}
\bigg\{1 + q\cdot
\Big[ {\cal S}_{f_{CP}} \sin(\Delta m_d \Delta t)
   + {\cal A}_{f_{CP}} \cos(\Delta m_d \Delta t)
\Big]
\bigg\},
\end{equation}
where $\tau_{B^0}$ is the $B^0$ lifetime, $\Delta m_d$ is
the mass difference between the two $B^0$ mass
eigenstates, $\Delta{t}$ = $t_{CP}$ $-$ $t_{\rm tag}$, and
the $b$-flavor charge $q$ = +1 ($-1$) when the tagging $B$ meson
is a $B^0$ ($\bar{B}^0$).
The $CP$-violating parameters ${\cal S}_{f_{CP}}$ and
${\cal A}_{f_{CP}}$  are given by
\begin{equation}
{\cal S}_{f_{CP}} \equiv \frac{2{\cal I}m(\lambda)}{|\lambda|^2+1}, \qquad
{\cal A}_{f_{CP}} \equiv \frac{|\lambda|^2-1}{|\lambda|^2+1},
\end{equation}
where $\lambda$ is a complex
parameter that depends on both the $B^0$-$\overline{B^0}$
mixing and on the amplitudes for $B^0$ and $\bar{B}^0$ decay to $f_{CP}$.
To a good approximation in the SM,
$|\lambda|$ is equal to the absolute value
of the ratio of the $\bar{B}^0 \rightarrow f_{CP}$ to
$B^0 \rightarrow f_{CP}$ decay amplitudes.

%DRM So far, in neutral $B$ meson decays mediated by the
%DRM $b \to c \bar{c} s$ transition,
%DRM $CP$ violation has been established by the measurements of
$CP$ violation in neutral $B$ meson decays involving the
$b \to c \bar{c} s$ transition has been established through  measurements of
the $CP$-violation  parameter $\sin2\phi_1$ by
the Belle~\cite{phi1_Belle} and BaBar~\cite{phi1_BaBar} collaborations.
Here, the SM predicts ${\cal S}_{f_{CP}} = -\xi_f \sin 2\phi_1$,
where $\xi_f = +1 (-1)$
corresponds to  $CP$-even (-odd) final states; and ${\cal A}_{f_{CP}} =0$
(or equivalently $|\lambda| = 1$) for both $b \rightarrow c\overline{c}s$
and
the leading contributions to $b \rightarrow c\bar{c}d$ 
(e.g. the tree diagram).
Hence, for ${f_{CP}}=J/\psi ~\pi^0$, which is a $CP$-even final state,
${\cal S}_{J/\psi\pi^0}$ becomes
$-\sin 2\phi_1$ if the tree diagram dominates.

%If penguin contributions or other contributions are 
%significant, a precision measurement of the
%time-dependent $CP$ asymmetry in $b \rightarrow c \bar{c} d$ may reveal
%values for ${\cal S}_{J/\psi\pi^0}$ and ${\cal A}_{J/\psi\pi^0}$ that
%differ from what is expected.   
If penguin contributions or other contributions are 
substantial, a precision measurement of the
time-dependent $CP$ asymmetry in $b \rightarrow c \bar{c} d$ may reveal
values for ${\cal S}_{J/\psi\pi^0}$ and ${\cal A}_{J/\psi\pi^0}$ that
differ from what is expected.   
Measurements of $CP$ asymmetries  in
$b \rightarrow c \bar{c} d$ transition-induced $B$ decays such as 
%DRM $B^0 \rightarrow J/\psi~\pi^0$ play an important role in ascertaining
$B^0 \rightarrow J/\psi~\pi^0$ thus play an important role in ascertaining
whether or not the KM model provides a complete description
of $CP$ violation in $B$ decays.

A study of $CP$ asymmetry in $B^0 \rightarrow J/\psi~\pi^0$ decays
has been reported by the BaBar collaboration\cite{psipi0CP_BaBar}.
In this paper
we report a measurement of time-dependent $CP$ violating parameters
in $B^0 \rightarrow J/\psi~\pi^0$ decays using the higher statistics data
accumulated by the Belle detector.

%%%%%
\section{Data sample and event selection}
%%%%%
The results presented here are based on a data sample of 140 fb$^{-1}$
(corresponding to 15.2$\times 10^7$ $B\bar{B}$ pairs) collected at
the $\Upsilon(4S)$ resonance with the Belle detector~\cite{Belle}
at the KEKB asymmetric-energy $e^+e^-$ (3.5 on 8~GeV) collider~\cite{KEKB}.
At KEKB, the $\Upsilon(4S)$ is produced
with a Lorentz boost of $\beta\gamma=0.425$ nearly along
the electron beamline ($z$).
%DRM Since the $B^0$ and $\bar{B}^0$ mesons are approximately at
Since the $B^0$ and $\bar{B}^0$ mesons are nearly at 
rest in the $\Upsilon(4S)$ center-of-mass system (cms),
$\Delta t$ can be determined from the displacement in $z$
between the $f_{CP}$ and $f_{\rm tag}$ decay vertices:
$\Delta t \simeq (z_{CP} - z_{\rm tag})/\beta\gamma c
\equiv \Delta z/\beta\gamma c$.

The Belle detector is a large-solid-angle magnetic
spectrometer that
consists of a three-layer silicon vertex detector (SVD),
a 50-layer central drift chamber (CDC), an array of
%DRM aerogel threshold \v{C}erenkov counters (ACC),
aerogel threshold Cherenkov counters (ACC),
a barrel-like arrangement of time-of-flight
scintillation counters (TOF), and an electromagnetic calorimeter
comprised of CsI(T${\it l}$) crystals (ECL) located inside
%DRM a super-conducting solenoid coil that provides a 1.5~T
a super-conducting solenoid coil that provides a 1.5-T
magnetic field.  An iron flux-return located outside of
the coil is instrumented to detect $K_L^0$ mesons and to identify
muons (KLM).  The detector
is described in detail elsewhere~\cite{Belle}.

%%%%%%%
% event selection
%%%%%%%
Hadronic events are selected if they satisfy the following criteria:
at least three reconstructed charged tracks;
a total reconstructed ECL energy in the center of mass (cms) frame
in the range between 0.1 and 0.8 times the total cms energy;
an average ECL cluster energy below 1~GeV;
at least one ECL shower in the region $-0.7 < \cos\theta < 0.9$ in
the laboratory frame;
a total visible energy, which is the sum of charged track momenta
and total ECL energy, exceeding 0.2 times the total cms energy;
and a reconstructed primary vertex that is consistent with the known
interaction point.
%DRM After the application of all these criteria,
After the imposition of these requirements, 
the efficiency for selecting $B$-meson pairs that include
%DRM a $J/\psi$ meson is estimated to be 99\%
%DRM by Monte Carlo simulation (MC).
a $J/\psi$ meson is estimated by Monte Carlo (MC) simulation to be 99\%.
To suppress continuum events, we require the event shape variable $R_2$
to be less than 0.5,
where $R_2$ is the ratio of the second to the zeroth
Fox-Wolfram moment~\cite{FWM}.

%-----------
%\subsection{Reconstruction of $J/\psi$ mesons}
%----------
$J/\psi$ mesons are reconstructed via their decay into oppositely 
charged lepton pairs ($e^+e^-$ or $\mu^+\mu^-$).   
Leptons are selected by starting with 
charged tracks satisfying $|dz|<5\mbox{~cm}$, where $dz$ is the track's 
closest approach to the interaction point along the beam direction.
For electron identification, the ratio between the charged track's
momentum and the associated shower energy ($E/p$) is the most powerful
discriminant.
Other information including $dE/dx$, the distance between the ECL shower
and the extrapolated track, and the shower shape are also used.
Muons are identified by requiring an association between KLM hits and
an extrapolated track.
%DRM Both lepton tracks have to be positively identified as such.
Both lepton tracks must be positively identified as such.
In the $e^+e^-$ mode, ECL clusters
that are within 50~mrad of the track's initial momentum vector are
included in the calculation of the invariant mass ($M_{ee(\gamma)}$),
in order to include photons radiated
from electrons/positrons.
The invariant masses of $e^+e^-(\gamma)$ and $\mu^+\mu^-$ combinations 
%DRM are required to fall in the range
are required to fall in the ranges  
$-0.15 < (M_{J/\psi} - M_{ee(\gamma)}) < +0.036\mbox{~GeV}/c^2$ and
$-0.06 < (M_{J/\psi} - M_{\mu\mu}) < +0.036\mbox{~GeV}/c^2$, respectively.
Here $M_{J/\psi}$ denotes the world average of the 
$J/\psi$ mass~\cite{PDG2003}.
%KM We perform a vertex fit to the lepton pair candidate.
%KM Then a mass constrained fit is applied
%KM to improve the $\Delta E$ resolution of the selected $B$ meson candidates.

%----------
%\subsection{Reconstruction of $\pi^0$ mesons}
%----------
Photon candidates are selected from clusters of up to 5$\times$5 crystals
in the ECL.
Each photon candidate is required to have no associated charged track,
and a cluster shape that is consistent with an electromagnetic shower.
%
%KM $\pi^0 \rightarrow \gamma \gamma$ decay candidates are formed from
%KM photon pairs that have an invariant mass
%KM in the range 0.118 to 0.15~GeV/$c^2$.
%KM The $\pi^0$ momentum is obtained by applying a mass constrained fit.
%KM To select $\pi^0$'s for the $B^0 \rightarrow J/\psi~\pi^0$ mode,
%KM the energy of each photon is required to exceed 50~MeV (100~MeV)
%KM in the ECL barrel (forward and backward endcap).
To select $\pi^0 \rightarrow \gamma \gamma$ decay candidates
for the $B^0 \rightarrow J/\psi~\pi^0$ mode,
the energy of each photon is required to exceed 50~MeV (100~MeV)
in the ECL barrel (forward and backward endcap).
$\pi^0$'s are formed from photon pairs that have an invariant mass
in the range 0.118 to 0.15~GeV/$c^2$.
%KM The $\pi^0$ momentum is obtained by applying a mass constrained fit.

\begin{figure}[htb]
\includegraphics[width=1.0\textwidth]{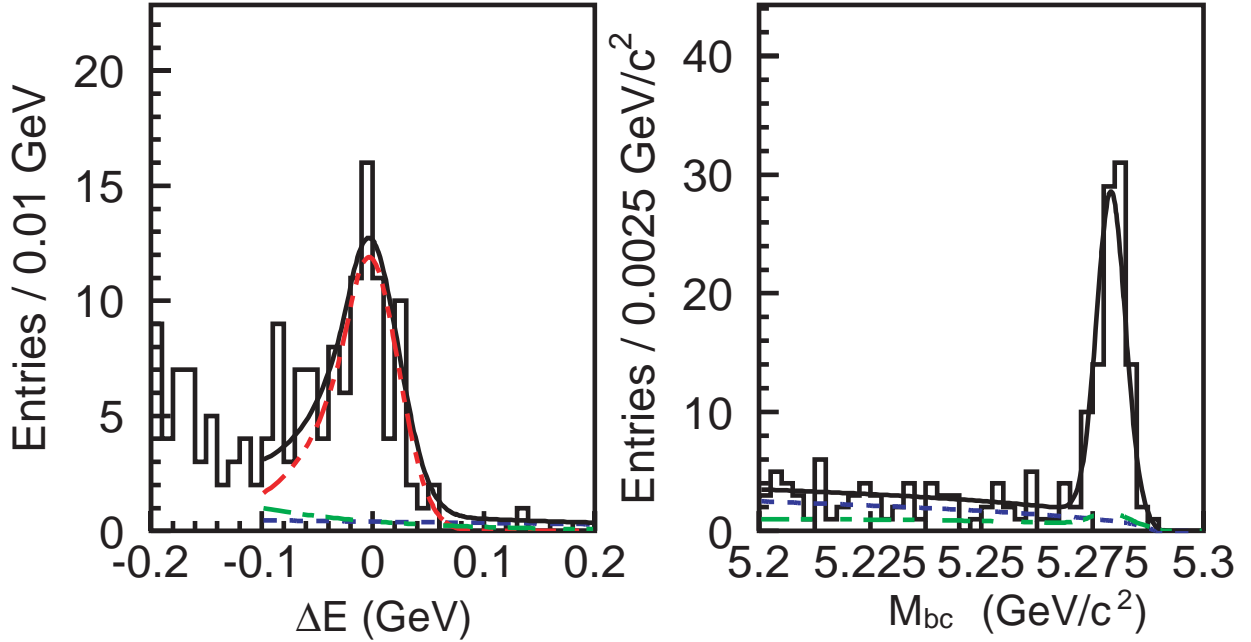}
\caption{The $\Delta E$(left) and $M_{\rm bc}$(right) distributions for
$B^0 \rightarrow J/\psi ~\pi^0$ candidates.
%%DRM I am not sure that I understand what the following
%%DRM sentence was trying to say.  Please check the 
%%DRM suggested replacement carefully to make sure that
%%DRM I have not changed the meaning.  
%DRM Superimposed curves shows fit result to obtain an event-by-event
%DRM signal probability; signal(plotted only in $\Delta E$
%DRM distribution), $B \rightarrow J/\psi X$ background, combinatorial
%DRM background and sum of all the contributions are drown by
%DRM red, green, gray dash and blue lines(See text).}
%KM The superimposed curves show fitted contributions from
%KM signal (red, plotted only in $\Delta E$
%KM distribution), $B \rightarrow J/\psi X$ background (green), 
%KM combinatorial background (gray dash) and the sum of all the 
%KM contributions (blue).  See text for further details.}
The superimposed curves show fitted contributions from
signal (red two-dot-dash, plotted only in $\Delta E$
distribution), $B \rightarrow J/\psi X$ background (green dot-dash), 
combinatorial background (blue dash) and the sum of all the 
contributions (black solid).  See text for further details.}

\label{mbc-deltae}
\end{figure}

$J/\psi$ and $\pi^0$ candidates are combined to select $B$ candidates.
The $B$ candidate selection is carried out using two observables
in the rest frame of the $\Upsilon(4S)$ (cms):
the beam-energy constrained mass
$M_{\rm bc} \equiv \sqrt{ E_{\rm beam}^2 - (\sum \vec{p_i})^2} $
and the energy difference $\Delta E \equiv \sum E_i - E_{\rm beam}$,
where $E_{\rm beam}=\sqrt{s}/2$ is the cms beam energy,
and $\vec{p_i}$ and $E_i$ are the cms three-momenta and energies of
the $B$ meson decay products.
%KM
In this calculation, kinematic fits are performed with
(1) vertex and mass constraints for the $J/\psi$ di-lepton decays
and (2) mass constraint for the $\pi^0 \rightarrow \gamma \gamma$
decays in order to improve the $\Delta E$ resolution.
%KM of the selected $B$ meson candidates.
%KM
Candidate events are selected by requiring
$5.270 < M_{\rm bc} < 5.290\mbox{~GeV}/c^2$
and $-0.10 < \Delta E < 0.05\mbox{~GeV}$.
%KM The lower limit of the $\Delta E$ requirement is used to 
%KM accommodate the negative $\Delta E$ tail that results from
%KM shower leakage associated with the high-momentum $\pi^0$.
The lower bound of the $\Delta E$ requirement is determined in order to 
accommodate the negative $\Delta E$ tail that results from
shower leakage associated with the high-momentum $\pi^0$.
%Note that with the loosened $\Delta E$ lower limit, we get more
%background events from
%$B^0 \rightarrow J/\psi ~K_S(K_S \rightarrow \pi^0 \pi^0)$.
The number of reconstructed $B^0 \rightarrow J/\psi ~\pi^0$ candidates is
103. The $\Delta E $ and $M_{\rm bc}$ distributions for the candidate events
are shown in Fig,~\ref{mbc-deltae}.
%--------- DeltaE and Mbc plots provided by KaiFeng ------
%----------------------------------------------------------

%%%%%
\section{Signal probability}
%%%%%
To assign an event-by-event signal probability for use in
the maximum-likelihood fit to the $CP$-violating parameters,
we determine event distribution functions in the $\Delta E$-$M_{\rm bc}$ plane
for both signal and background.
The signal distribution is
modeled with a two-dimensional function which is 
Gaussian in $M_{\rm bc}$ and uses a Crystal Ball line shape~\cite{CBlineshape}
in $\Delta E$.  The shape parameters of these functions are determined 
%DRM from Monte Carlo (MC) simulation and the signal yield is allowed to float.
from MC simulation and held fixed in the fit, 
while the overall signal yield is allowed to float.
Backgrounds are studied using a large sample of MC events along
with events outside of the signal region.  
%DRM We split the backgrounds into two categories, one is $B$ decays having
%DRM a $J/\psi$($B \rightarrow J/\psi X$) and the other is combinatorial
We split the backgrounds into two categories, one being $B$ decays having
a $J/\psi$($B \rightarrow J/\psi X$) and the other being combinatorial
background to which random combinations of particles in
$B\bar{B}$ decays and continuum events contribute.
According to MC study, the $B \rightarrow J/\psi X$ background forms
a small peak in the $M_{\rm bc}$ projection.
Therefore we parametrize this contribution with the sum of a 
Gaussian and
a phase-space like background function(ARGUS function)~\cite{ARGUSBG}
in the $M_{\rm bc}$ direction and an exponential function for $\Delta E$.
The amount of this background contribution is determined by
MC~\cite{inclusivepsi}.
For the combinatorial background, we use a linear function
for $\Delta E$ and an ARGUS function for $M_{\rm bc}$.
The purity of the signal is estimated to be 86$\pm$10\%

%%%%
\section{Flavor tagging and vertexing}
%%%%
Charged leptons, kaons, pions, and $\Lambda$ baryons
that are not associated with the reconstructed
$B^0 \rightarrow J/\psi ~\pi^0$ decay
are used to identify the $b$-flavor of the accompanying $B$ meson,
denoted by $f_{\rm tag}$.
Based on the measured properties of these tracks, two parameters,
$q$ and $r$, are assigned to each event.
The first, $q$, has the discrete value $+1$~($-1$)
when the tag-side $B$ meson is more likely to be a $B^0$~($\bar{B}^0$).
The parameter $r$ is an event-by-event MC-determined
flavor-tagging dilution factor that ranges
from $r=0$ for no flavor discrimination
to $r=1$ for an unambiguous flavor assignment.
It is used only to sort data into six intervals of $r$,
according to the estimated flavor purity.
%KM The wrong-tag probabilities for each of these intervals, $w_l~(l=1,6)$,
%KM which are used in the final fit, are determined directly from the data.
%KM The difference of the wrong-tag fractions between $B^0$ and $\bar{B}^0$
%KM %DRM are concerned and implemented as 
%KM %DRM $w^{dif}_l(l=1,6)$ into the final fit.
%KM are also determined from data and implemented as $w^{dif}_l(l=1,6)$ 
%KM into the final fit.
%KM Samples of $B^0$ decays to exclusively reconstructed self-tagging channels
%KM are utilized to obtain $w_l$ and $w^{dif}_l$ using
%KM time-dependent $B^0$-$\bar{B}^0$ mixing:
%KM $(N_{\rm OF}-N_{\rm SF})/(N_{\rm OF}+N_{\rm SF}) =
%KM (1-2w_l)\cos(\Delta m\Delta t)$,
%KM where $N_{\rm OF}$ and $N_{\rm SF}$ are the numbers of opposite
%KM ($B^0\bar{B}^0 \rightarrow B^0\bar{B}^0$) and
%KM same ($B^0\bar{B}^0 \rightarrow B^0B^0, \bar{B}^0\bar{B}^0$) flavor events.
The wrong-tag probabilities for each of these intervals,$w_l~(l=1,6)$, 
which are used in the final fit, are determined directly from the data 
samples of $B^0$ decays to exclusively reconstructed self-tagging channels.
The difference of the wrong-tag fractions between $B^0$ and $\bar{B}^0$
are also determined from the same samples 
and implemented as $\Delta w_l(l=1,6)$ into the final fit.
We obtain $w_l$ and $\Delta w_l$ using 
time-dependent $B^0$-$\bar{B}^0$ mixing:
$(N_{\rm OF}-N_{\rm SF})/(N_{\rm OF}+N_{\rm SF}) =
(1-2w_l)\cos(\Delta m\Delta t)$,
where $N_{\rm OF}$ and $N_{\rm SF}$ are the numbers of opposite
($B^0\bar{B}^0 \rightarrow B^0\bar{B}^0$) and
same ($B^0\bar{B}^0 \rightarrow B^0B^0, \bar{B}^0\bar{B}^0$) flavor events.
%The event fractions and wrong tag fractions for each $r$ interval
%are described elsewhere~\cite{Belle_sin2phi1_2003}.
The wrong tag fractions for each $r$ interval
are given elsewhere~\cite{Belle_sin2phi1_2003}.

%%%%%
% vertexing.
%%%%%
The decay vertices of $B^0$ mesons are reconstructed using
tracks that have enough SVD hits: i.e. both $z$ and $r$-$\phi$
hits in at least one SVD layer and at least one additional
layer with a $z$ hit, where the $r$-$\phi$ plane is
perpendicular to the $z$ (beams) axis.
Each vertex position is required to be consistent with
the IP profile, which is determined run-by-run and
smeared in the $r$-$\phi$ plane by 21~$\mu$m
to account for the $B$ meson decay length.
%KM With these requirements, we are able to determine a vertex
%KM even with a single track.
With these requirements, we are able to determine a vertex even
in the case where only one track has enough associated SVD hits.
The vertex position for the $B^0 \rightarrow J/\psi ~\pi^0$ decay
is reconstructed using lepton tracks from the $J/\psi$.
The algorithm for
the $f_{\rm tag}$ vertex reconstruction is chosen to minimize
the effect of long-lived particles, secondary vertices from
charmed hadrons and a small fraction of poorly reconstructed
tracks~\cite{resol_nim}.
From all the charged tracks with associated SVD hits
except those used for $B^0 \rightarrow J/\psi ~\pi^0$ reconstruction,
we select tracks with a position error in the $z$ direction of
less than 500 $\mu$m, and with an impact parameter with respect to
the $J/\psi$ vertex of less than 500 $\mu$m.
Track pairs with opposite charges are removed
if they have an invariant mass
within $\pm15$~MeV/$c^2$ of the nominal $K_S$ mass.
If the reduced $\chi^2$ associated with 
the $f_{\rm tag}$ vertex exceeds 20,
the track making the largest $\chi^2$ contribution is removed and
the vertex is refitted.
This procedure is repeated until an acceptable 
reduced $\chi^2$ is obtained.
After flavor tagging and vertex reconstruction, we obtain 91
$B^0 \rightarrow J/\psi ~\pi^0$ candidates.

%%%%%
\section{The unbinned maximum likelihood fit}
%%%%
We determine ${\cal S}_{J/\psi \pi^0}$ and ${\cal A}_{J/\psi \pi^0}$
for each mode by performing
an unbinned maximum-likelihood fit to the observed $\Delta t$ distribution.
%The probability density function (PDF) expected for the signal
%distribution is given by Eq.~(\ref{eq:psig}) with $q$ replaced by
%$q(1-2w_l-\Delta w_l)$ to account for the effect of incorrect flavor
%assignment.
The probability density function (PDF) expected for the signal
distribution is given by 
\begin{eqnarray}
\lefteqn{{\cal P}_{\rm sig}(\Delta t,q,w_l,\Delta w_l)} \nonumber \\
&=& 
\frac{e^{-|\Delta_{t}|/{\tau_{B^0}}}}{4{\tau_{B^0}}}
\bigg\{1 -q\Delta w_l + q(1-2w_l)\cdot
\Big[ {\cal S}_{f_{J/\psi \pi^0}} \sin(\Delta m_d \Delta t)
   + {\cal A}_{f_{J/\psi \pi^0}} \cos(\Delta m_d \Delta t)
\Big]
\bigg\}
\end{eqnarray}
to account for the effect of incorrect flavor
assignment.
The distribution is
convolved with the
proper-time interval resolution function $R_{\rm sig}(\Delta t)$,
which takes into account the finite vertex resolution.
$R_{\rm sig}(\Delta t)$ is formed by convolving four components: 
the detector resolutions for $z_{CP}$ and
%DRM $z{\rm tag}$, the shift in the $z_{\rm tag}$ vertex position
$z_{\rm tag}$, the shift in the $z_{\rm tag}$ vertex position
due to secondary tracks originating from charmed particle decays, and
the kinematic approximation that the $B$ mesons are at rest in the
cms~\cite{resol_nim}.
A small component of broad outliers in the $\Delta z$ distribution, caused
by
mis-reconstruction, is represented by a Gaussian function
$P_{\rm ol}(\Delta t)$.
We determine twelve resolution parameters and the neutral- and
charged-$B$ lifetimes simultaneously
from a fit to the $\Delta t$ distributions of hadronic $B$ decays and
obtain an average $\Delta t$ resolution of $\sim 1.43~$ps (rms).
We determine the following likelihood value for each
event:
\begin{eqnarray}
\lefteqn{P_i(\Delta t_i;{\cal S}_{J/\psi\pi^0},{\cal A}_{J/\psi\pi^0})}
\nonumber \\
&=& (1-f_{\rm ol})\int_{-\infty}^{\infty}\biggl[
f_{\rm sig}{\cal P}_{\rm sig}(\Delta t',q,w_l,\Delta w_l)R_{\rm sig}
(\Delta t_i-\Delta t')  \nonumber \\
& & \quad + f_{\rm bkg}^{J/\psi X}{\cal P}_{\rm bkg}^{J/\psi X}(\Delta t')
R_{\rm bkg}^{J/\psi X}(\Delta t_i-\Delta t') \nonumber \\
& &\quad +\; (1-f_{\rm sig}-f_{\rm bkg}^{J/\psi X})
{\cal P}_{\rm bkg}^{\rm comb}(\Delta t')
R_{\rm bkg}^{\rm comb}(\Delta t_i-\Delta t')\biggr]
d(\Delta t') + f_{\rm ol} P_{\rm ol}(\Delta t_i)
\end{eqnarray}
where $f_{\rm ol}$ is the outlier fraction and
$f_{\rm sig}$ is the signal probability calculated as a function
of $\Delta E$ and $M_{\rm bc}$.
${\cal P}_{\rm bkg}^{J/\psi X}(\Delta t)$ and
${\cal P}_{\rm bkg}^{\rm comb}(\Delta t)$
are  the PDFs for $B \rightarrow J/\psi X$ and combinatorial background events,
respectively.
These contributions dilute the significance of $CP$ violation in
Eq.\ (\ref{eq:psig}).
They are modeled as a sum of exponential and prompt components, and
are convolved with the corresponding resolution functions
$R_{\rm bkg}^{J/\psi X}$ and $R_{\rm bkg}^{\rm comb}$, respectively.
The resolution functions are modeled by a sum of two Gaussians.
All parameters in ${\cal P}_{\rm bkg}^{J/\psi X} (\Delta t)$
and $R_{\rm bkg}^{J/\psi X}$ are determined from MC simulation,
%%DRM I didn't understand what was meant by ``because its amount is fixed''?
%KM Your question is reasonable, amount is also one of the determined 
%KM parameters by MC. The corrected expression is right.
while the parameters in
${\cal P}_{\rm bkg}^{\rm comb} (\Delta t)$
and $R_{\rm bkg}^{\rm comb}$ are determined by a fit to
the $\Delta t$ distribution
of a background-enhanced control sample;
i.e. events away from the $\Delta E$-$M_{\rm bc}$ signal region.
We fix  $\tau_{B^0}$ and $\Delta m_d$ at
their world-average values~\cite{PDG2003}.
The only free parameters in the final fit
are ${\cal S}_{J/\psi \pi^0}$ and ${\cal A}_{J/\psi \pi^0}$,
which are determined by maximizing the
likelihood function
\begin{equation}
{\cal L} = \prod_iP_i(\Delta t_i;{\cal S}_{J/\psi\pi^0},{\cal
A}_{J/\psi\pi^0})
\end{equation}
where the product is over all events.

%%%%%%%%
% Results
%DRM \section{The $CP$-violation parameters}
\section{FIT RESULTS} 
\begin{figure}[htb]
\includegraphics[width=0.7\textwidth]{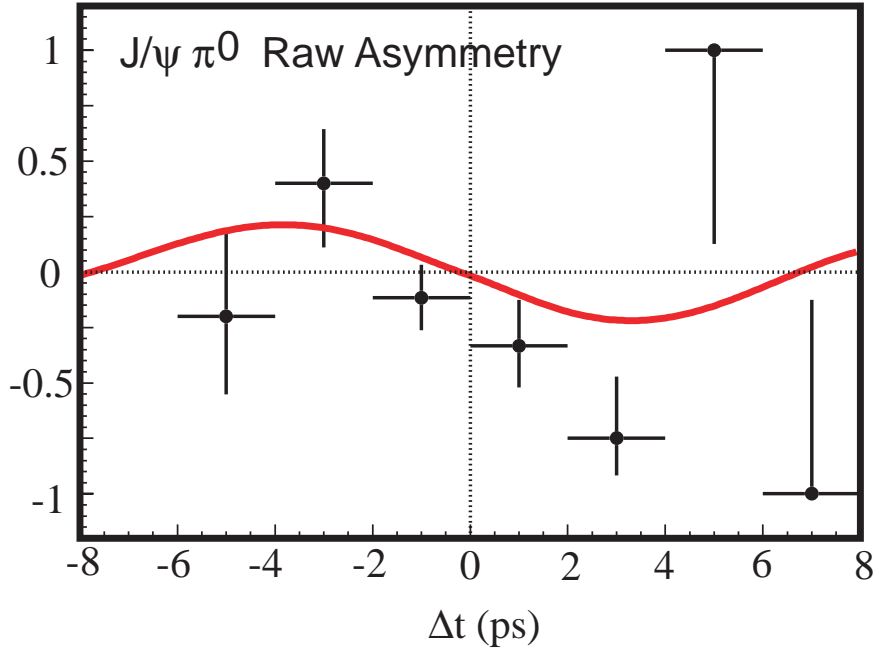}
\caption{The $\Delta t$ asymmetry. The curve shows the result
of the unbinned-maximum likelihood fit.}
\label{raw_asym}
\end{figure}

%%%%%%%%
A fit to the candidate events results in the $CP$-violation parameters;
%\begin{eqnarray}\nonumber
%{\cal S}_{J/\psi \pi^0}
%&=& -0.72 \displaystyle{^{+0.42}_{-0.37}}(\mbox{stat})
%\pm x.xx(\mbox{syst}) \\
%{\cal A}_{J/\psi \pi^0}
%&=& -0.01 \displaystyle{^{+0.29}_{-0.28}}(\mbox{stat})
%\pm x.xx(\mbox{syst}).  
%\end{eqnarray}
\begin{eqnarray}\nonumber
{\cal S}_{J/\psi \pi^0}
&=& -0.72 \pm 0.42(\mbox{stat}) \pm 0.08(\mbox{syst}) \\
{\cal A}_{J/\psi \pi^0}
&=& -0.01 \pm 0.29(\mbox{stat}) \pm 0.07(\mbox{syst}),  
\end{eqnarray}
%%DRM next line added.  
where the sources of systematic error are given below.
Figure \ref{deltat} shows the $\Delta t$ distributions for 
$\bar{B}^0 \rightarrow J/\psi~\pi^0$ (upper figure:$q=+1$) and
$B^0 \rightarrow J/\psi~\pi^0$ (lower figure:$q=-1$) event samples.
%--------- Delta t plot goes here ------
\begin{figure}[htb]
\includegraphics[width=0.5\textwidth]{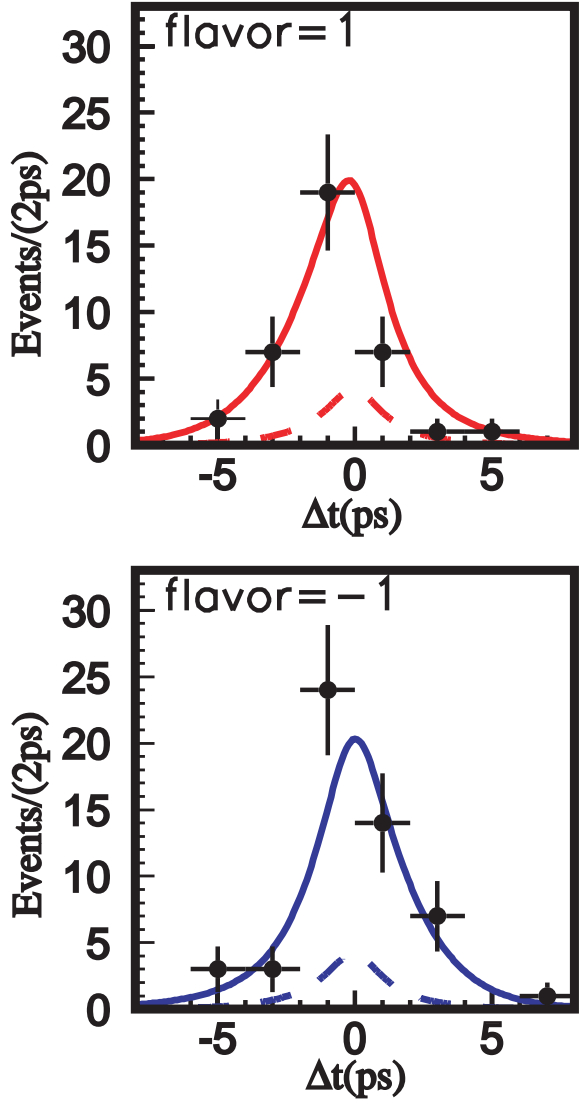}
\caption{The $\Delta t$ distributions for
$\bar{B}^0 \rightarrow J/\psi~\pi^0$ (upper:$q=+1$) and
$B^0 \rightarrow J/\psi~\pi^0$ (lower:$q=-1$) candidates.
The solid curves show the results of the global fits,
and dashed curves show the background distributions.}
\label{deltat}
\end{figure}
%----------------------------------------------------------
%
Figure \ref{raw_asym} shows the raw asymmetry in each $\Delta t $ bin
without background subtraction, which is defined by 
\begin{eqnarray}
A \equiv { { N_{q=+1} - N_{q=-1}  }\over{ N_{q=+1} + N_{q=-1} } }
\end{eqnarray}
where $N_{q=+1}$($N_{q=-1}$) is the number of observed candidates
with $q=+1$$(-1)$. The curve shows the result of unbinned-maximum 
likelihood fit to the $\Delta t$ distribution, 
${\cal S}_{J/\psi \pi^0}\sin(\Delta m_d \Delta t) + 
{\cal A}_{J/\psi \pi^0}\cos(\Delta m_d \Delta t)$.
%--------- Asymmetry plot goes here ------
%----------------------------------------------------------
%%DRM the following point is made in the ``Summary and Conclusions''
%%DRM section near the end.
%DRM Since ${\cal A}_{J/\psi \pi^0}$ is consistent to zero and 
%DRM ${\cal S}_{J/\psi \pi^0}$ coincide with $-\sin 2\phi_1$, 
%DRM our results favor that the SM tree diagram dominates in 
%DRM $B^0(\bar{B}^0) \rightarrow J/\psi ~|pi^0$ decays.

%DRM the following section heading was added.
\section{SYSTEMATIC UNCERTAINTIES}
%KM We consider systematic uncertainty as follows;
We estimate systematic uncertainties as follows;
the flavor tagging 
($\pm 0.025$ for ${\cal S}_{J/\psi \pi^0}$ and
$\pm 0.014$ for ${\cal A}_{J/\psi \pi^0}$),  
the signal probability 
($\pm 0.023$ for ${\cal S}_{J/\psi \pi^0}$ and
$\pm 0.016$ for ${\cal A}_{J/\psi \pi^0}$),  
the background $\Delta t$ distribution
($\pm 0.014$ for ${\cal S}_{J/\psi \pi^0}$ and
$\pm 0.0065$ for ${\cal A}_{J/\psi \pi^0}$),  
the resolution function
($\pm 0.010$ for ${\cal S}_{J/\psi \pi^0}$ and
$\pm 0.0061$ for ${\cal A}_{J/\psi \pi^0}$),  
the potential fit bias 
($\pm 0.043$ for ${\cal S}_{J/\psi \pi^0}$ and
$\pm 0.059$ for ${\cal A}_{J/\psi \pi^0}$),  
the vertex reconstruction 
($\pm 0.062$ for ${\cal S}_{J/\psi \pi^0}$ and
$\pm 0.018$ for ${\cal A}_{J/\psi \pi^0}$),  
and $B$ meson's lifetime and mixing parameter
($\pm 0.0019$ for ${\cal S}_{J/\psi \pi^0}$ and
$\pm 0.0071$ for ${\cal A}_{J/\psi \pi^0}$).
The quadratic sum of all the contribution mentioned above
amounts $\pm 0.0845$ for ${\cal S}_{J/\psi \pi^0}$ and
$\pm 0.0662$ for ${\cal A}_{J/\psi \pi^0}$.

%%%%
%DRM \section{Summary}
\section{SUMMARY AND CONCLUSIONS} 
%%%%
%DRM In summary, we have performed measurement of $CP$ violation parameters
We have performed a measurement of $CP$-violation parameters
in $B^0 \rightarrow J/\psi ~\pi^0$ decay.
%DRM The resultant $CP$ violation parameters are 
The resultant values are 
${\cal S}_{J/\psi \pi^0}
= -0.72 \pm 0.42(\mbox{stat}) \pm 0.08(\mbox{syst})$
and 
${\cal A}_{J/\psi \pi^0}
= -0.01 \pm 0.29(\mbox{stat}) \pm 0.07(\mbox{syst})$.
%DRM They are consistent with those
These values are consistent with those
obtained for $B^0 \rightarrow J/\psi ~K_S$ and other decays governed
by $b \rightarrow c\bar{c}s$ transition and suggest that penguin
%DRM contribution to this decay mode is not large.
and other contributions to this decay mode are not large.

%
%\begin{table}[htb]
%\caption{ This is an example of a table.}
%\label{sys1}
%\begin{tabular}
%{@{\hspace{0.5cm}}l@{\hspace{0.5cm}}||@{\hspace{0.5cm}}c@{\hspace{0.5cm}}}
%\hline \hline
%Source & Systematic error (\%) \\
%\hline
%ISR correction & $\pm 19$ \\
%Fitting procedure & $\pm 16$ \\
%$J/\psi$ polarization & $\pm 11$ \\
%Track reconstruction  & $\pm 5$ \\
%Lepton identification  & $\pm 4$ \\
%\hline
%Total & $\pm 28$ \\
%\hline \hline
%\end{tabular}
%\end{table}
%
%{\bf See http://belle.kek.jp/$\sim$kinosh/private/pub/figure\_tips.html for
%advice and macros for publication quality figures.
%Files for figures should be collected together in a single directory.
%The filenames for figures should be correspond to the numbering in the
%paper e.g. psiks\_fig1.eps, psiks\_fig2.eps, psiks\_fig3.eps. These
%figures will be made available to conference speakers and reviewers.
%The text of the paper should refer to all relevant
%Belle publications. For example, relevant conference papers
%should refer to previous Belle papers on CP violation\cite{psiks_Belle},
%\cite{pipi_Belle}.}
%
\section*{Acknowledgments}
% Please paste this acknowledgment into your latex file.
%***** Acknowledgments *****
We wish to thank the KEKB accelerator group for the excellent
operation of the KEKB accelerator.
We acknowledge support from the Ministry of Education,
Culture, Sports, Science, and Technology of Japan
and the Japan Society for the Promotion of Science;
the Australian Research Council
and the Australian Department of Industry, Science and Resources;
the National Science Foundation of China under contract No.~10175071;
the Department of Science and Technology of India;
the BK21 program of the Ministry of Education of Korea
and the CHEP SRC program of the Korea Science and Engineering
Foundation;
the Polish State Committee for Scientific Research
under contract No.~2P03B 01324;
the Ministry of Science and Technology of the Russian Federation;
the Ministry of Education, Science and Sport of the Republic of
Slovenia;
the National Science Council and the Ministry of Education of Taiwan;
and the U.S.\ Department of Energy.

\end{document}

%% file: author-conf2003.tex
%%% Paper:    (fill in)
%%% Journal:  Summer 2003 conference proceedings (Physical Review format)
%%% Contacts: (fill in)
%%% Each author is included unless he/she chooses to opt out of ALL papers.
%%% ====================================================================
%%% Click the RELOAD button on your web browser to see the updated file.
%%% ====================================================================
%%% Use \input{author} to insert this material into your latex file.
%%%%% Force institutions to appear in alphabetical order when typeset.
\affiliation{Aomori University, Aomori}
\affiliation{Budker Institute of Nuclear Physics, Novosibirsk}
\affiliation{Chiba University, Chiba}
\affiliation{Chuo University, Tokyo}
\affiliation{University of Cincinnati, Cincinnati, Ohio 45221}
\affiliation{University of Frankfurt, Frankfurt}
\affiliation{Gyeongsang National University, Chinju}
\affiliation{University of Hawaii, Honolulu, Hawaii 96822}
\affiliation{High Energy Accelerator Research Organization (KEK), Tsukuba}
\affiliation{Hiroshima Institute of Technology, Hiroshima}
\affiliation{Institute of High Energy Physics, Chinese Academy of Sciences, Beijing}
\affiliation{Institute of High Energy Physics, Vienna}
\affiliation{Institute for Theoretical and Experimental Physics, Moscow}
\affiliation{J. Stefan Institute, Ljubljana}
\affiliation{Kanagawa University, Yokohama}
\affiliation{Korea University, Seoul}
\affiliation{Kyoto University, Kyoto}
\affiliation{Kyungpook National University, Taegu}
\affiliation{Institut de Physique des Hautes \'Energies, Universit\'e de Lausanne, Lausanne}
\affiliation{University of Ljubljana, Ljubljana}
\affiliation{University of Maribor, Maribor}
\affiliation{University of Melbourne, Victoria}
\affiliation{Nagoya University, Nagoya}
\affiliation{Nara Women's University, Nara}
\affiliation{National Kaohsiung Normal University, Kaohsiung}
\affiliation{National Lien-Ho Institute of Technology, Miao Li}
\affiliation{Department of Physics, National Taiwan University, Taipei}
\affiliation{H. Niewodniczanski Institute of Nuclear Physics, Krakow}
\affiliation{Nihon Dental College, Niigata}
\affiliation{Niigata University, Niigata}
\affiliation{Osaka City University, Osaka}
\affiliation{Osaka University, Osaka}
\affiliation{Panjab University, Chandigarh}
\affiliation{Peking University, Beijing}
\affiliation{Princeton University, Princeton, New Jersey 08545}
\affiliation{RIKEN BNL Research Center, Upton, New York 11973}
\affiliation{Saga University, Saga}
\affiliation{University of Science and Technology of China, Hefei}
\affiliation{Seoul National University, Seoul}
\affiliation{Sungkyunkwan University, Suwon}
\affiliation{University of Sydney, Sydney NSW}
\affiliation{Tata Institute of Fundamental Research, Bombay}
\affiliation{Toho University, Funabashi}
\affiliation{Tohoku Gakuin University, Tagajo}
\affiliation{Tohoku University, Sendai}
\affiliation{Department of Physics, University of Tokyo, Tokyo}
\affiliation{Tokyo Institute of Technology, Tokyo}
\affiliation{Tokyo Metropolitan University, Tokyo}
\affiliation{Tokyo University of Agriculture and Technology, Tokyo}
\affiliation{Toyama National College of Maritime Technology, Toyama}
\affiliation{University of Tsukuba, Tsukuba}
\affiliation{Utkal University, Bhubaneswer}
\affiliation{Virginia Polytechnic Institute and State University, Blacksburg, Virginia 24061}
\affiliation{Yokkaichi University, Yokkaichi}
\affiliation{Yonsei University, Seoul}
  \author{K.~Abe}\affiliation{High Energy Accelerator Research Organization (KEK), Tsukuba} % KEK
  \author{K.~Abe}\affiliation{Tohoku Gakuin University, Tagajo} % TohokuGakuin
  \author{N.~Abe}\affiliation{Tokyo Institute of Technology, Tokyo} % TIT
  \author{R.~Abe}\affiliation{Niigata University, Niigata} % Niigata
  \author{T.~Abe}\affiliation{High Energy Accelerator Research Organization (KEK), Tsukuba} % KEK
  \author{I.~Adachi}\affiliation{High Energy Accelerator Research Organization (KEK), Tsukuba} % KEK
  \author{Byoung~Sup~Ahn}\affiliation{Korea University, Seoul} % Korea
  \author{H.~Aihara}\affiliation{Department of Physics, University of Tokyo, Tokyo} % Tokyo
  \author{M.~Akatsu}\affiliation{Nagoya University, Nagoya} % Nagoya
  \author{M.~Asai}\affiliation{Hiroshima Institute of Technology, Hiroshima} % Hiroshima
% ...
% Lausanne
\collaboration{The Belle Collaboration}

%% file: bf0342.bbl
\begin{thebibliography}{99}
\bibitem{CC}
Throughout this paper,
the inclusion of the charge conjugate mode decay is implied
unless otherwise stated.
\bibitem{CBlineshape}
A description of this function can be found in
Ian C. Brock, BONN-MS-99-02, pp86.
\bibitem{inclusivepsi}
Two-body $B$ meson decays including a $J/\psi$ meson, whose
branching fractions are well-measured($B \rightarrow J/\psi K^{(*)}$),
is dominant contribution.
According to MC, $B^0 \rightarrow J/\psi ~K_S$ and 
$B^0 \rightarrow J/\psi ~K_L$ decays contribute equally to 
the peaking background in the $M_{\rm bc}$ projection.
They have opposite $CP$ eigenvalues and thus their $CP$ violating
contributions cancel.
\end{thebibliography}
